\documentclass[aps,prr,reprint,amsmath,amssymb,groupedaddress]{revtex4-2}

\usepackage{graphicx}% Include figure files
\usepackage{dcolumn}% Align table columns on decimal point
\usepackage{bm}% bold math
\usepackage{color}
\usepackage{ulem}
\usepackage{comment}
\newcommand{\ua}{\uparrow}
\newcommand{\da}{\downarrow}
\newcommand{\dg}{\dagger}

\begin{document}

%\preprint{}

\title{Unified description of cuprate superconductors using four-band $d$-$p$ model}


\author{Hiroshi Watanabe$^{1,2}$}
\email{h-watanb@fc.ritsumei.ac.jp}
\author{Tomonori Shirakawa$^{3,4}$}
\author{Kazuhiro Seki$^5$}
\author{Hirofumi Sakakibara$^{2,6,7}$}
\author{Takao Kotani$^{6,7}$}
\author{Hiroaki Ikeda$^8$}
\author{Seiji Yunoki$^{2,3,4,5}$}
\affiliation{
$^1$Research Organization of Science and Technology, Ritsumeikan University, Shiga 525-8577, Japan\\
$^2$Computational Condensed Matter Physics Laboratory, RIKEN Cluster for Pioneering Research (CPR), Saitama 351-0198, Japan\\
$^3$Computational Materials Science Research Team, RIKEN Center for Computational Science (R-CCS), Hyogo 650-0047, Japan\\
$^4$Quantum Computational Science Research Team, RIKEN Center for Quantum Computing (RQC), Saitama 351-0198, Japan\\
$^5$Computational Quantum Matter Research Team, RIKEN Center for Emergent Matter Science (CEMS), Saitama 351-0198, Japan\\
$^6$Advanced Mechanical and Electronic System Research Center (AMES), Faculty of Engineering, Tottori University, Tottori 680-8552, Japan\\
$^7$Center of Spintronics Research Network (CSRN), Graduate School of Engineering Science, Osaka University, Osaka, 560-8531, Japan\\
$^8$Department of Physics, Ritsumeikan University, Shiga 525-8577, Japan
}
%\thanks{}


\date{\today}

\begin{abstract}
In the 35 years since the discovery of cuprate superconductors, we have not yet reached a unified understanding of their properties, including their material dependence of the superconducting transition temperature $T_{\text{c}}$.
The preceding theoretical and experimental studies have provided an overall picture of the phase diagram, and some important parameters for the $T_{\text{c}}$,
such as the contribution of the Cu $d_{z^2}$ orbital to the Fermi surface and the site-energy difference $\Delta_{dp}$ between the Cu $d_{x^2-y^2}$ and O $p$ orbitals.
However, they are somewhat empirical and limited in scope, always including exceptions, and do not provide a comprehensive view of the series of cuprates.
Here we propose a four-band $d$-$p$ model as a minimal model to study material dependence in cuprates.
Using the variational Monte Carlo method, we theoretically investigate the phase diagram for the La$_2$CuO$_4$ and HgBa$_2$CuO$_4$ systems and the correlation between the key parameters and the superconductivity.
Our results comprehensively account for the empirical correlation between $T_{\text{c}}$ and model parameters, and thus can provide a guideline for new material design.
We also show that the effect of the nearest-neighbor $d$-$d$ Coulomb interaction $V_{dd}$ is actually quite important for the stability of superconductivity and phase competition.
\end{abstract}

%\keywords{}

\maketitle

\section{INTRODUCTION}\label{intro}
The discovery of superconductivity in cuprates has brought about the significant progress in strongly-correlated electron systems~\cite{Bednorz}.
Along with the high superconducting transition temperature $T_{\text{c}}$, cuprates show anomalous phases and phenomena such as the Mott metal-insulator transition, pseudogap phenomena, and 
antiferromagnetic (AF) and charge-density-wave (or stripe) phases~\cite{Imada,Uchida}.
Recently, nematic order~\cite{Sato} and ferromagnetic fluctuation~\cite{Sonier,Kurashima} have also been observed experimentally.
These newly observed experiments being constantly reported with the underlying, possibly exotic, physics have continued to attract many researchers' interests over 35 years since the first high-$T_{\text{c}}$ cuprate superconductor was synthesized.

It has been widely believed that the strong AF spin fluctuation characteristic to the two-dimensional square lattice structure triggers the various anomalous phenomena in cuprates~\cite{Scalapino,Miyake,Kampf,Moriya}.
On the basis of this picture, the effective one-band Hubbard model~\cite{Anderson} has been studied with various numerical methods extensively~\cite{LeBlanc,Zheng}.
It succeeded to describe the important physics in cuprates not only for the ground state but also for the excited state.
The angle-resolved photoemission spectroscopy (ARPES) experiments show that the Fermi surface consists of only one energy band mainly derived from the Cu $d_{x^2-y^2}$ orbital~\cite{Damascelli} and thus the one-band models are justified as long as the low-energy physics is concerned.

However, recent theoretical and experimental development sheds light on the importance of the orbital degree of freedom in cuprates.
For example, material dependence of $T_{\text{c}}$ is difficult to discuss within the one-band models.
The previous studies of the one-band models generally discussed the material dependence via the shape of different Fermi surfaces originated by different hopping integrals, e.g., the nearest-neighbor hopping $t$ and the next nearest-neighbor hopping $t'$.
Indeed, such treatment allowed us to capture the overall feature for the difference between hole-doped and electron-doped systems~\cite{Yanase}.
However, the material dependence of $T_{\text{c}}$ for the hole-doped systems has not been able to be explained properly.
Experimentally, $T_{\text{c}}$ becomes higher with larger $|t'/t|$, i.e., more rounded Fermi surface~\cite{Pavarini}.
On the contrary, it is predicted theoretically that in most one-band models, $T_{\text{c}}$ is reduced because of poor nesting properties in rounded Fermi surfaces~\cite{Maier,Sakakibara}.
A clue to resolving this contradiction can be found by seriously considering the orbital degrees of freedom, especially, the Cu $d_{z^2}$ orbital~\cite{Sakakibara,Uebelacker,Miyahara}.
The study for the two-orbital model shows that the contribution of the $d_{z^2}$ orbital to the Fermi surface, which is strong in low $T_{\text{c}}$ materials, works against the $d_{x^2-y^2}$-wave superconductivity~\cite{Sakakibara}.
This picture can explain the material dependence of $T_{\text{c}}$.
Indeed, a recent ARPES experiment directly observed the hybridization gap between the $d_{z^2}$ and $d_{x^2-y^2}$ orbitals and pointed out the importance of the multiorbital effect on superconductivity~\cite{Matt}.

Another extension to a multiorbital model is the introduction of the O $p$ orbitals.
A minimal model that includes the Cu $d_{x^2-y^2}$, O $p_x$, and O $p_y$ orbitals in the CuO$_2$ plane is known as the (three-band) $d$-$p$ model or the Emery model~\cite{Emery}.
The three-band $d$-$p$ model has been studied with various numerical methods~\cite{Asahata,Takimoto,Yanagisawa,Lorenzana,Shinkai,Thomale,Kent,Weber1,Weber2,Weber3,Fischer,Weber4,Bulut,Yamakawa,Ogura,White,Tsuchiizu,Orth,Zegrodnik,Dash,Moreo,Biborski,Cui,Chiciak,Mai} as much as the one-band models.
In these studies, the site-energy difference $\Delta_{dp}$ between the Cu $d_{x^2-y^2}$ and O $p_{x/y}$ orbitals is found to be an important parameter for understanding the material dependence of $T_{\text{c}}$.
For example, the three-band $d$-$p$ model can successfully reproduce the negative correlation between $\Delta_{dp}$ and $T_{\text{c}}$~\cite{Weber4}, namely, a smaller $\Delta_{dp}$ leads to a higher $T_{\text{c}}$.
Furthermore, the role of the O $p$ orbitals in the superconducting pairing~\cite{Ogura,Zegrodnik,Moreo,Biborski,Mai}, stripe order~\cite{Lorenzana,White}, loop current~\cite{Thomale,Weber1,Fischer}, and nematic order~\cite{Fischer,Yamakawa,Tsuchiizu,Orth}
has been discussed with the three-band $d$-$p$ model.
However, it is still difficult to correctly capture the correlation between Fermi surface topology and $T_{\text{c}}$ within the three-band $d$-$p$ model~\cite{Kent,Weber4}.

In this paper, we study a four-band $d$-$p$ model composed of the Cu $d_{x^2-y^2}$ and $d_{z^2}$ orbitals and the O $p_x$ and $p_y$ orbitals, which can be expected to resolve the problems mentioned above.
As typical examples of single-layer hole-doped cuprates, we consider La$_2$CuO$_4$ and HgBa$_2$CuO$_4$ systems.
We construct the tight-binding model for these systems based on the first-principles calculation and examine the effect of Coulomb interaction by the variational Monte Carlo (VMC) method.
We show that the material dependence of $T_{\text{c}}$ is well explained with two key parameters, the site energy $\varepsilon_{d_{z^2}}$ of the Cu $d_{z^2}$ orbital and the site-energy difference $\Delta_{dp}$ between the Cu $d_{x^2-y^2}$ and O $p_{x/y}$ orbitals,
which is consistent with the empirical relation.
We thus propose that the present four-band $d$-$p$ model is a minimal model that can properly describe the material dependence of cuprate superconductors.
Furthermore, we also study the effect of the nearest-neighbor $d$-$d$ Coulomb interaction $V_{dd}$, which has not been discussed in detail previously.
We show that $V_{dd}$ substantially affects the stability of superconductivity and the phase competition among various phases, suggesting an important parameter for the effective model of cuprates.
In addition, we find that $V_{dd}$ induces a crossover from a Slater insulator to a Mott insulator at the undoped limit.

The rest of this paper is organized as follows.
In Sec.~\ref{model}, a four-band $d$-$p$ model on the two-dimensional square lattice is introduced.
The VMC method and the variational wave functions are also explained in Sec.~\ref{model}.
The numerical results are then provided in Sec.~\ref{results}.
The tight-binding energy bands for the La$_2$CuO$_4$ and HgBa$_2$CuO$_4$ systems, obtained on the basis of the first-principles calculation, are first shown in Sec.~\ref{band}.
The material and doping dependences of a superconducting correlation function are then examined and the role of two key parameters $\varepsilon_{d_{z^2}}$ and $\Delta_{dp}$ are clarified in Sec.~\ref{SC}.
The effect of the nearest-neighbor $d$-$d$ Coulomb interaction $V_{dd}$ is investigated in Sec.~\ref{Vdd}.
The phase competition among superconductivity and other phases is briefly discussed in Sec.~\ref{competition}.
Finally, the paper concludes with a summary in Sec.~\ref{summary}.
The details of the variational wave functions are described in Appendix.

\section{MODEL AND METHOD}\label{model}
\subsection{Four-band $d$-$p$ model}\label{d-p}
\begin{figure}
\centering
\includegraphics[width=0.9\hsize]{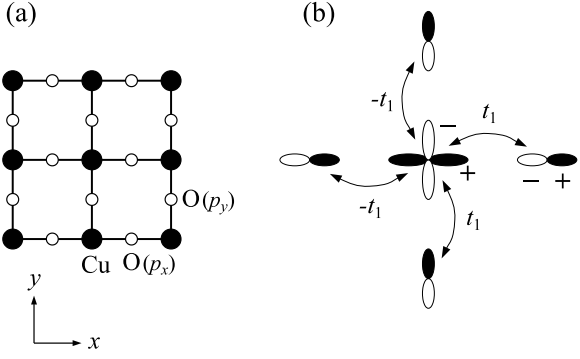}
\caption{\label{lattice}
(a) Schematic lattice structure of the four-band $d$-$p$ model, forming the two-dimensional square lattice.
Each Cu site contains $d_{x^2-y^2}$ and $d_{z^2}$ orbitals, while there is either $p_x$ or $p_y$ orbital on O sites.
The lattice constant between the nearest-neighbor Cu sites is set to be one, i.e., primitive translation vectors $|\textbf{e}_x|=|\textbf{e}_y|=1$.
(b) Phase convention for the Cu $d_{x^2-y^2}$ orbital and the O $p_x$ and $p_y$ orbitals.
Solid (open) ovals indicate the positive (negative) phase.
The hopping integral between the $d_{x^2-y^2}$ and $p_{x/y}$ orbitals, $t_1$, is shown with the sign.
The definitions of other hopping integrals ($t_2-t_6$) are described in Appendix~\ref{NIband}.
}
\end{figure}

As an effective low-energy model of cuprates, we consider a four-band $d$-$p$ model on the two-dimensional square lattice (see Fig.~\ref{lattice}) defined by the following Hamiltonian:
\begin{equation}
H=H_{\text{kin}}+H_{\text{int}}-H_{\text{dc}}.
\end{equation} 
Here, the kinetic term $H_{\text{kin}}$ is described by
\begin{align}
H_{\text{kin}}&=\sum_{i,j,\sigma}\sum_{\alpha,\beta}t^{\alpha\beta}_{ij}c^{\dg}_{i\alpha\sigma}c_{j\beta\sigma} \label{kin_orb} \\
                &=\sum_{\textbf{k},\sigma}\sum_m E_m(\textbf{k})a^{\dg}_{\textbf{k}m\sigma}a_{\textbf{k}m\sigma}, \label{kin_band}
\end{align}
where Eq.~(\ref{kin_orb}) is the kinetic term in an orbital representation and Eq.~(\ref{kin_band}) is in a band representation.
$c^{\dg}_{i\alpha\sigma}$ ($c_{i\alpha\sigma}$) is a creation (annihilation) operator of an electron at site $i$ with spin $\sigma\,(=\ua,\da)$ and orbital $\alpha\, (=1,2,3,4)$ 
corresponding to ($d_{x^2-y^2}$, $d_{z^2}$, $p_x$, $p_y$), respectively.
$t^{\alpha\beta}_{ij}$ denotes a hopping integral between orbital $\alpha$ at site $i$ and orbital $\beta$ at site $j$.
$t^{\alpha\alpha}_{ii}$ is a site energy $\varepsilon_{\alpha}$ for orbital $\alpha$ at site $i$.
Equation~(\ref{kin_band}) is obtained by diagonalizing Eq.~(\ref{kin_orb}), and the energy eigenvalue $E_m(\textbf{k})$ is characterized by the wave vector $\textbf{k}$ and the energy band index $m\,(=1,2,3,4)$.
$a^{\dg}_{\textbf{k}m\sigma}$ ($a_{\textbf{k}m\sigma}$) is a creation (annihilation) operator of the corresponding energy band with spin $\sigma$. 
The undoped parent compounds of cuprates correspond to one hole (i.e., seven electrons) per unit cell in this model, which is conventionally referred to as half filling, and hereafter we denote the carrier density as the number $\delta$ of holes per unit cell 
that are introduced into the system at half filling.

\begin{figure}
\centering
\includegraphics[width=0.7\hsize]{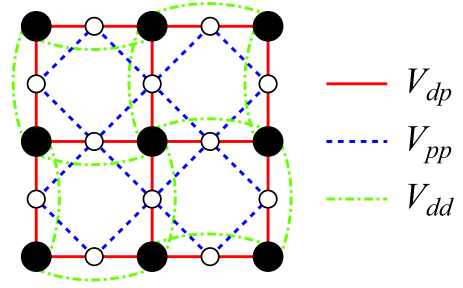}
\caption{\label{interaction}
The Coulomb interaction parameters between nearest-neighbor orbitals.
Solid (open) circles represent Cu (O) atoms.
}
\end{figure}

The Coulomb interaction term $H_{\text{int}}$ is composed of eight terms,
\begin{align}
H_{\text{int}}&=U_d\sum_i\left(n^{d_1}_{i\ua}n^{d_1}_{i\da}+n^{d_2}_{i\ua}n^{d_2}_{i\da}\right) \notag \\
 &+\left(U'_d-\frac{J}{2}\right)\sum_in^{d_1}_in^{d_2}_i-2J\sum_i\textbf{S}^{d_1}_i\cdot\textbf{S}^{d_2}_i \notag \\
 &-J'\sum_i\left(c^{\dg}_{i1\ua}c^{\dg}_{i1\da}c_{i2\ua}c_{i2\da}+c^{\dg}_{i2\ua}c^{\dg}_{i2\da}c_{i1\ua}c_{i1\da}\right) \notag \\
 &+U_p\sum_i\left(n^{p_x}_{i\ua}n^{p_x}_{i\da}+n^{p_y}_{i\ua}n^{p_y}_{i\da}\right)+V_{dp}\sum_{\left<i,j\right>}n^d_in^{p_{x/y}}_j \notag \\
 &+V_{pp}\sum_{\left<i,j\right>}n^{p_x}_in^{p_y}_j+V_{dd}\sum_{\left<i,j\right>}n^d_in^d_j. \label{int}
\end{align}
Here, $n^{\alpha}_i=n^{\alpha}_{i\ua}+n^{\alpha}_{i\da}$ with $n^{\alpha}_{i\sigma}=c^{\dg}_{i\alpha\sigma}c_{i\alpha\sigma}$ is the number operator and $\textbf{S}^{\alpha}_i$ is the spin angular momentum operator at site $i$ with orbital $\alpha$.
$d_1$ and $d_2$ are abbreviations for $d_{x^2-y^2}$ and $d_{z^2}$ orbitals, respectively, and $n^d_i=n^{d_1}_i+n^{d_2}_i$.
$U_d,U_d',J,$ and $J'$ represent on-site intraorbital, interorbital, Hund's coupling, and pair-hopping interactions between $d$ orbitals, respectively.
In this study, we set $J'=J$ and $U_d=U'_d+2J$~\cite{Kanamori}.
The on-site intraorbital Coulomb interaction within $p$ orbitals, $U_p$, is also introduced.
The last three terms in $H_{\text{int}}$ take into account the intersite Coulomb interactions between nearest-neighbor orbitals, $V_{dp},V_{pp},$ and $V_{dd}$, as shown in Fig.~\ref{interaction},
where the sum $\sum_{\left<i,j\right>}$ runs over all pairs of nearest-neighbor orbitals located at site $i$ and $j$.

In addition, the following double counting correction term $H_{\text{dc}}$ is introduced,
\begin{multline}
H_{\text{dc}}=\biggl[\left\{U_d+2\left(U'_d-\frac{J}{2}\right)+16V_{dd}\right\}\langle n^d\rangle_0\ \\
\shoveright{+8V_{dp}\left<n^p\right>_0\biggr]\sum_in^d_i} \\
+\left\{(U_p+8V_{pp})\left<n^p\right>_0+8V_{dp}\langle n^d\rangle_0\right\}\sum_i (n^{p_x}_i+n^{p_y}_i),
\end{multline}
where $\langle n^d\rangle_0=\frac{1}{N_{\text{S}}}\sum_i\langle n^d_i\rangle$ and $\left<n^p\right>_0=\frac{1}{N_{\text{S}}}\sum_i\langle (n^{p_x}_i+n^{p_y}_i)\rangle$ are the average electron density of the $d$ and $p$ orbitals in the noninteracting limit
and $N_{\text{S}}$ is the total number of unit cells.
When we apply a many-body calculation method to a multiorbital model, the site energy of each orbital is shifted due to the interaction effect.
However, such energy shifts have already been included in the energy band of the tight-binding model constructed from the first-principles calculation.
This is a so-called double counting problem and should be treated with care especially in the $d$-$p$ model~\cite{Hansmann}.
Here, we subtract the term $H_{\text{dc}}$ from the Hamiltonian to correct the energy shift.
This is one of the reasonable treatments to avoid the double counting.

\begin{table*}
\centering
\caption{\label{hopping}
The tight-binding parameters for the La$_2$CuO$_4$ and HgBa$_2$CuO$_4$ systems in eV units estimated on the basis of maximally localized Wannier orbitals from the first-principles LDA calculation.
For comparison, the tight-binding parameters for the La$_2$CuO$_4$ system with reference to the QSGW method are also shown (denoted as ``revised'').
The definitions of $t_i$ and $\varepsilon_{\alpha}$ are described in Appendix~\ref{NIband}.
$\Delta_{dp}=\varepsilon_{d_{x^2-y^2}}-\varepsilon_{p}$, i.e., the site-energy difference between the Cu $d_{x^2-y^2}$ and O $p_{x/y}$ orbitals.
}
\begin{tabular}{lccccccccccc}
\hline
\hline
 & $t_1$ & $t_2$ & $t_3$ & $t_4$ & $t_5$ & $t_6$ & $\varepsilon_{d_{x^2-y^2}}$ & $\varepsilon_{d_{z^2}}$ & $\varepsilon_{p_{x/y}}$ & $\Delta_{dp}$ \\
La$_2$CuO$_4$   & 1.42 & 0.61 & 0.07 & 0.65 & 0.05 & 0.07 & -0.87 & -0.11 & -3.13 & 2.26\\
La$_2$CuO$_4$(revised) & 1.42 & 0.61 & 0.07 & 0.51 & 0.03 & 0.07 & -0.87 & -0.68 & -3.13 & 2.26\\
HgBa$_2$CuO$_4$ & 1.26 & 0.65 & 0.13 & 0.33 & 0.00 & 0.05 & -1.41 & -1.68 & -3.25 & 1.84\\
\hline
\end{tabular}
\end{table*}

\begin{figure*}
\centering
\includegraphics[width=1.0\hsize]{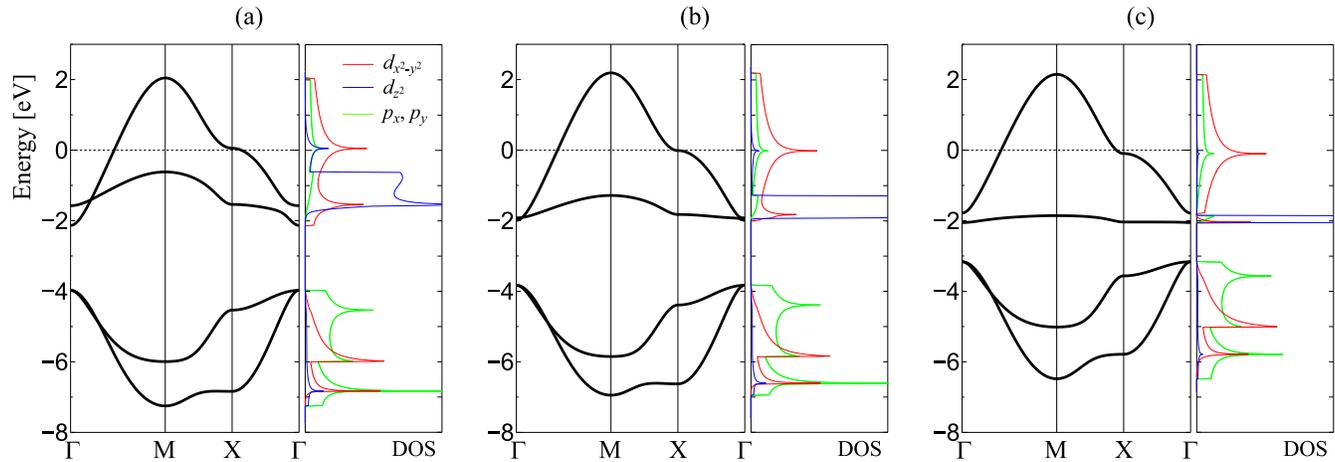}
\caption{\label{band-DOS}
The energy dispersions and projected density of states onto the $d_{x^2-y^2}$, $d_{z^2}$, and $p_{x/y}$ orbitals for the noninteracting tight-binding models for (a) La$_2$CuO$_4$, (b) La$_2$CuO$_4$(revised), and (c) HgBa$_2$CuO$_4$ systems.
The tight-binding parameters are given in Table~\ref{hopping}.
Fermi energy (defined as zero energy) is set to be the case of 15\% hole doping ($\delta$=0.15).
The high symmetric momenta are indicated as $\Gamma=(0,0)$, $\text{M}=(\pi,\pi)$, and $\text{X}=(\pi,0)$.
}
\end{figure*}

\subsection{VMC method}\label{VMC}
The effect of Coulomb interaction is treated using a VMC method~\cite{McMillan,Ceperley,Yokoyama1}.
The trial wave function considered here is a Gutzwiller-Jastrow type composed of four parts,
\begin{equation}
\left|\Psi\right>=P^{(2)}_{\text{G}}P_{\text{J}_{\text{c}}}P_{\text{J}_{\text{s}}}\left|\Phi\right>.
\label{twf}
\end{equation}
$\left|\Phi\right>$ is a one-body part constructed by diagonalizing the one-body Hamiltonian including the off-diagonal elements 
$\{\rho^{\alpha}_i\}$, $\{m^{\alpha}_i\}$, and $\{\Delta^{\alpha\beta}_{ij}\}$, which induce long-range ordering of charge, spin, and superconductivity, respectively.
The renormalized hopping integrals $\{\tilde{t}^{\alpha\beta}_{ij}\}$ are also included in $\left|\Phi\right>$ as variational parameters.
The explicit forms of them are described in Appendix.

The Gutzwiller factor
\begin{equation}
P^{(2)}_{\text{G}}=\prod_{i,\gamma}\bigl[\text{e}^{-g_{\gamma}}
\left|\gamma\right>\left<\gamma\right|_i\bigr]
\end{equation}
is extended to two-orbital systems~\cite{Bunemann,Watanabe1}.
In $P^{(2)}_\text{G}$, possible 16 patterns of charge and spin configuration of the $d_{x^2-y^2}$ and $d_{z^2}$ orbitals at each site $\left| \gamma \right>$, i.e.,
$\left|0\right>=\left|0\;0\right>$, $\left|1\right>=\left|0\ua\right>$, $\cdots$, $\left|15\right>=\left|\ua\da\;\ua\da\right>$,
are differently weighted with $\text{e}^{-g_{\gamma}}$ and $\{g_{\gamma}\}$ are optimized as variational parameters.

The remaining operators
\begin{equation}
P_{\text{J}_{\text{c}}}=\exp\Bigl[-\sum_{i,j}\sum_{\alpha,\beta}v^{\text{c}}_{ij\alpha\beta}n^{\alpha}_in^{\beta}_j\Bigr]
\end{equation}
and
\begin{equation}
P_{\text{J}_{\text{s}}}=\exp\Bigl[-\sum_{i,j}\sum_{\alpha,\beta}v^{\text{s}}_{ij\alpha\beta}s^z_{i\alpha}s^z_{j\beta}\Bigr]
\end{equation}
are charge and spin Jastrow factors, which control long-range charge and spin correlations, respectively.
$s^z_{i\alpha}$ is the $z$ component of the spin angular momentum operator at site $i$ with orbital $\alpha$.
We set $v^{\text{c}}_{ii\alpha\beta}=v^{\text{s}}_{ii\alpha\beta}=0$ for $\alpha,\beta=d_1,d_2$ because the on-site correlation between the $d_{x^2-y^2}$ and $d_{z^2}$ orbitals are already taken into account in $P^{(2)}_{\text{G}}$.

In this paper, we focus mainly on the superconducting correlation functions to examine where the superconductivity appears in the phase diagram.
The detailed studies on other competing orders will be discussed elsewhere.
The variational parameters in $\left|\Psi\right>$ are therefore
$\{\tilde{t}^{\alpha\beta}_{ij}\}$, $\{\Delta^{\alpha\beta}_{ij}\}$, $\{g_{\gamma}\}$, $\{v^{\text{c}}_{ij\alpha\beta}\}$, and $\{v^{\text{s}}_{ij\alpha\beta}\}$.
They are simultaneously optimized using stochastic reconfiguration method~\cite{Sorella}.
We show results for $N_{\text{S}}$=24$\times$24=576 unit cells (and thus 576$\times$4=2304 orbitals in total), which is large enough to avoid finite size effects.
The antiperiodic boundary conditions are set for both $x$ and $y$ directions of the primitive lattice vectors.

\section{RESULTS}\label{results}
\subsection{Band structures of La$_2$CuO$_4$ and HgBa$_2$CuO$_4$}\label{band}
First, we discuss the material dependence of the band structure.
As a typical example of single-layer hole-doped cuprates, we study the La$_2$CuO$_4$ and HgBa$_2$CuO$_4$ systems.
We construct maximally localized Wannier orbitals~\cite{Marzari,Souza} from the first-principles calculation in the local-density approximation (LDA) with ecalj package~\cite{ecalj}
and fit them with the hopping integrals $t_i$ ($i=1-6$) and the site energy of each orbital $\varepsilon_{\alpha}$.
The parameter sets determined for these systems are listed in Table~\ref{hopping} and the explicit form of the tight-binding model is described in Appendix~\ref{NIband}.
Note that the estimated site energy of the Cu $d_{z^2}$ orbital, $\varepsilon_{d_{z^2}}$, and the hybridization between the Cu $d_{z^2}$ and O $p_{x/y}$ orbitals, $t_4$, depend significantly 
on the method of first-principles calculation.
In fact, the estimated $\varepsilon_{d_{z^2}}$ is much lower in the quasiparticle self-consistent $GW$ (QSGW) method~\cite{Faleev,vanShilfgaarde,Kotani3} than in the conventional LDA calculation~\cite{Jang1}.
To clarify the effect, we also consider another parameter set of La system with reference to the QSGW band structure (labeled as ``revised''),
where the site energy $\varepsilon_{d_{z^2}}$ is lower and $t_4$ is also slightly smaller than those estimated on the basis of the LDA calculation (See Table~\ref{hopping}).

Figure~\ref{band-DOS} shows the noninteracting tight-binding energy bands for La and Hg systems.
We can notice the clear difference among them:
(i) The density of states (DOS) of the $d_{z^2}$ component is extended from 0 to -2~eV in the La system [Fig.~\ref{band-DOS}(a)], while it is almost localized around -2~eV in the Hg system [Fig.~\ref{band-DOS}(c)].
This is because the $d_{z^2}$ orbital is hybridized with the $p_{x/y}$ orbital in the La system much more strongly than in the Hg system.
The $d_{z^2}$ electrons obtain the itinerancy through the hybridization and therefore the $d_{z^2}$ component of the La system is more extended than that of the Hg system.
The revised version of the La system is located somewhere in between [Fig.~\ref{band-DOS}(b)].
(ii) The site-energy difference between the $d_{x^2-y^2}$ and $p_{x/y}$ orbitals, $\Delta_{dp}=\varepsilon_{d_{x^2-y^2}}-\varepsilon_{p}$, is larger in the La system than in the Hg system.
This affects the occupancy of each orbital and thus the strength of the electron correlation.
For example, when $\Delta_{dp}\, (>0)$ is small, the energy band crossing the Fermi energy contains more component of the $p_{x/y}$ orbital, in which the intraorbital Coulomb interaction is smaller.

Starting from these energy band structures, we shall investigate the ground state property of the La and Hg systems using the VMC method.
We assume that the tight-binding parameters remain unchanged with hole doping.
The Coulomb interaction parameters are set as
$(U_d,U'_d,J,U_p,V_{dp},V_{pp})=(8.0, 6.4, 0.8, 4.0, 2.0, 1.6)\;t_1$ for both La and Hg systems with reference to Ref.~\cite{Hirayama}.
Note that the values for the La system are larger in eV units because of the larger $t_1$.
In the following, we set $t_1$ as a unit of energy.
We first set $V_{dd}=0$ and then discuss the effect of finite $V_{dd}$.
As shown in Sec.~\ref{Vdd}, we find that even small $V_{dd}$ can substantially affect the property of the system.

\subsection{Superconducting correlation function}\label{SC}
\subsubsection{Overview}
To discuss the material dependence of superconductivity, we calculate the superconducting correlation function defined as
\begin{equation}
P^{dd}(\textbf{r})=\frac{1}{N_{\text{S}}}\sum_i\sum_{\tau,\tau'}f^{(dd)}_{\tau\tau'}\bigl<\Delta^{\dg}_{\tau}(\textbf{R}_i) \Delta_{\tau'}(\textbf{R}_i+\textbf{r})\bigr>,
\end{equation}
where $\Delta^{\dg}_{\tau}(\textbf{R}_i)$ is a creation operator of singlet pairs between nearest-neighbor $d_{x^2-y^2}$ orbitals,
\begin{equation}\label{Delta}
\Delta^{\dg}_{\tau}(\textbf{R}_i)=\frac{1}{\sqrt{2}}(c^{\dg}_{i1\uparrow}c^{\dg}_{i+\tau1\downarrow}+c^{\dg}_{i+\tau1\uparrow}c^{\dg}_{i1\downarrow}),
\end{equation}
and $\tau$ runs over four nearest-neighbor Cu sites ($\tau=\pm \mathbf{e}_x,\pm\mathbf{e}_y$).
$f^{(dd)}_{\tau\tau'}$ is a form factor of a superconducting gap function with $d_{x^2-y^2}$ symmetry, namely,
$f^{(dd)}_{\tau\tau'}=1$ for $\tau \parallel \tau'$ and $-1$ for $\tau \perp \tau'$.
$\langle \cdots \rangle$ denotes $\langle\Psi|\cdots|\Psi\rangle / \langle\Psi|\Psi\rangle$ for the optimized variational wave function $|\Psi\rangle$.
If $P^{dd}(\textbf{r})$ is saturated to a finite value for $r=|\mathbf{r}|\rightarrow\infty$, superconducting long-range order exists.

\begin{figure}
\centering
\includegraphics[width=0.8\hsize]{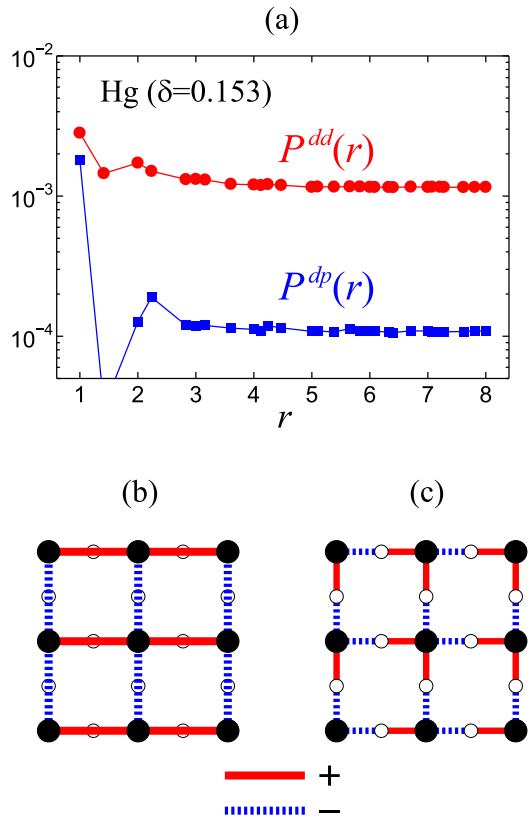}
\caption{\label{SC-1}
(a) Superconducting correlation function $P^{dd}(r)$ and $P^{dp}(r)$ for the Hg system at a hole doping rate $\delta=0.153$.
The sign change of (b) $d$-$d$ and (c) $d$-$p$ pairings.
Cu (O) sites are indicated by solid (open) circles in (b) and (c).
}
\end{figure}

Figure~\ref{SC-1}(a) shows the behavior of $P^{dd}(r=|\textbf{r}|)$ for the Hg system at a hole doping rate $\delta=0.153$.
It shows good convergence for $r \gtrsim 4$ and reveals that the superconducting long-range order certainly exists.
The sign of the $d$-$d$ pairing in a real space is shown in Fig.~\ref{SC-1}(b).
It is positive in the $x$ direction and negative in the $y$ direction, reflecting the $d_{x^2-y^2}$-wave symmetry expected in cuprate superconductors.
We also calculate the superconducting correlation function for pairing formed between the $d_{x^2-y^2}$ and $p_{x/y}$ orbitals $P^{dp}(r)$, which is defined in the same way as $P^{dd}(r)$
except that $c^{\dg}_{i+\tau1\sigma}$ in Eq.~(\ref{Delta}) is replaced with $c^{\dg}_{i+\frac{\tau}{2}3(4)\sigma}$ for $\tau=\pm\textbf{e}_{x(y)}$.
Although the value is one order of magnitude smaller than $P^{dd}(r)$ [see Fig.~\ref{SC-1}(a)], $P^{dp}(r)$ is also saturated to a finite value, indicative of the long-range order of $d$-$p$ pairing.
Note that the sign of the $d$-$p$ pairing changes alternatively along both $x$ and $y$ directions and thus shows $p$-like symmetry as shown in Fig.~\ref{SC-1}(c).
This is due to the phase convention of $p_x$ and $p_y$ orbitals adopted here [see Fig.~\ref{lattice}(b)] and is consistent with the $d_{x^2-y^2}$ pairing symmetry~\cite{Nomoto}.
It is also compatible with the preceding study of the three-band $d$-$p$ model~\cite{Cui}.
This kind of real space or orbital representation is useful for the analysis of cuprate superconductors~\cite{Moreo,Biborski,Mai} because the pairing length is expected to be very short~\cite{Li}.

\begin{figure}
\centering
\includegraphics[width=0.8\hsize]{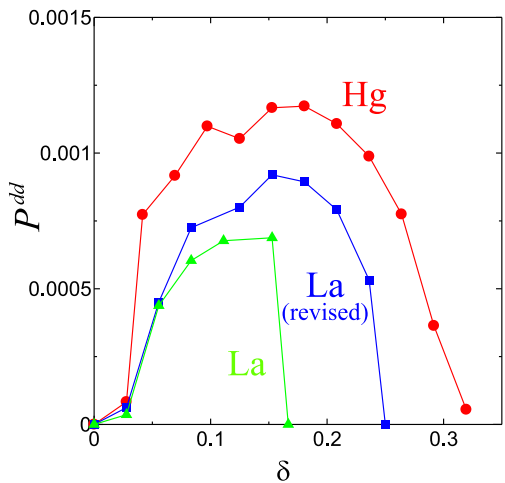}
\caption{\label{SC-2}
$P^{dd}$ as a function of the hole doping rate $\delta$ for the three systems.
}
\end{figure}

\subsubsection{Material dependence: Effect of $d_{z^2}$ orbital}
Let us now examine the material and doping dependence of the superconducting correlation function.
We take the converged value of $P^{dd}(r\rightarrow\infty)$ as a strength of superconductivity $P^{dd}$. 
Figure~\ref{SC-2} shows the doping dependence of $P^{dd}$ for the La, La(revised), and Hg systems.
For all cases, $P^{dd}$ displays a dome shape as a function of the hole doping rate $\delta$.
At $\delta=0$, the system is insulating due to the strong correlation effect and thus $P^{dd}=0$.
As $\delta$ increases, mobile carriers are introduced into the system and the mobility of the Cooper pair increases.
On the other hand, the strength of the $d$-$d$ pairing itself is reduced by doping because the electron correlation effect is also reduced.
The balance between these two factors results in the dome-shaped behavior of $P^{dd}$.
This picture is expected to be universal for all hole-doped cuprate superconductors.

\begin{figure}
\centering
\includegraphics[width=1.0\hsize]{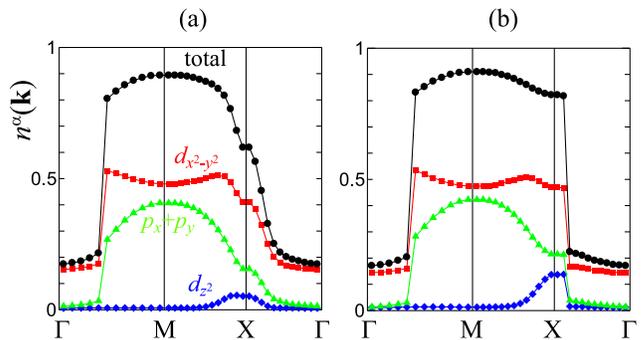}
\caption{\label{n(k)}
Total and $\alpha$ orbital components of momentum distribution function for the La system at (a) $\delta=0.153$ (superconducting phase) and (b) $\delta=0.181$ (paramagnetic metallic phase).
The high symmetric momenta are indicated as $\Gamma=(0,0)$, $\text{M}=(\pi,\pi)$, and $\text{X}=(\pi,0)$.
}
\end{figure}

Next, let us study the importance of the orbital character near the Fermi energy in the material dependence of superconductivity.
Generally, a large DOS at the Fermi energy is favorable for superconductivity due to the large energy gain of gap opening.
However, a detailed structure of the DOS, namely, the orbital character and $\mathbf{k}$ dependence should be carefully investigated.
For this purpose, we calculate the following momentum distribution function of holes,
\begin{equation}
n^{\alpha}(\textbf{k})=\frac{1}{2}\sum_{\sigma} \langle c_{\textbf{k}\alpha\sigma}c^{\dg}_{\textbf{k}\alpha\sigma} \rangle,
\end{equation}
where $c^{\dg}_{\textbf{k}\alpha\sigma}$ ($c_{\textbf{k}\alpha\sigma}$) is a Fourier transform of $c^{\dg}_{i\alpha\sigma}$ ($c_{i\alpha\sigma}$) in Eq.~(\ref{kin_orb}).
Figure~\ref{n(k)} shows $n^{\alpha}(\textbf{k})$ for the La system at the superconducting phase ($\delta=0.153$) and the paramagnetic metallic phase ($\delta=0.181$).
The discontinuities around the X point and in the middle of the $\Gamma$-M line in Fig.~\ref{n(k)}(b) for the paramagnetic metallic phase indicate the existence of Fermi surface.
Since the node of the superconducting gap runs along the $\Gamma$-M line, a discontinuity also exists in the superconducting phase, as shown in Fig.~\ref{n(k)}(a).
We can observe that the $d_{z^2}$ component has large weight around the X point, exhibiting a peak structure in $n^{\alpha}(\textbf{k})$, where the superconducting gap becomes the largest (so-called ``hot spot'').
This feature is expected to be unfavorable for superconductivity because the AF spin fluctuation, which promotes the $d_{x^2-y^2}$-wave pairing, is suppressed when the Cu $d_{z^2}$ orbital contributes to the formation of the Fermi surface~\cite{Kuroki}.

\begin{figure}
\centering
\includegraphics[width=0.7\hsize]{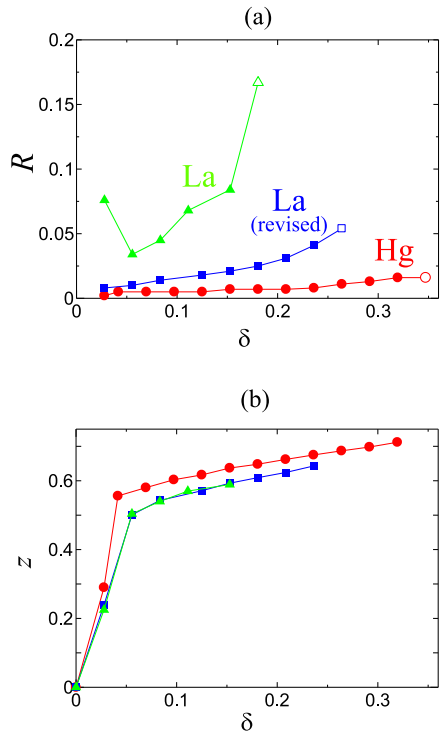}
\caption{\label{Rz}
(a) $\delta$ dependence of the ratio $R$ of the $d_{z^2}$ component to the total at the X point for the three systems.
Solid (open) symbols represent the superconducting (paramagnetic metallic) phase.
(b) $\delta$ dependence of the quasiparticle renormalization factor $z$ for the three systems.
Symbols are the same with those in (a).
}
\end{figure}

To investigate the effect of the $d_{z^2}$ component, the ratio of the $d_{z^2}$ component to the total at the X point, $R=n^{d_{z^2}}(\text{X})/n^{\text{tot}}(\text{X})$, is calculated in Fig~\ref{Rz}(a).
For the La system, the ratio $R$ increases with increasing $\delta$ and shows rapid enhancement for $\delta>0.16$.
It coincides with the sudden disappearance of $P^{dd}$ for $\delta>0.16$ shown in Fig.~\ref{SC-2}.
On the other hand, $R$ for the La(revised) system is much smaller than that of the La system, because $\varepsilon_{d_{z^2}}$ is lower and the hybridization between the $d_{z^2}$ and $d_{x^2-y^2}$ orbitals via the $p_{x/y}$ orbitals is smaller.
As a result, the superconducting phase is extended to a larger value of $\delta$ and a smooth dome shape is observed in $P^{dd}$ vs. $\delta$ as shown in Fig.~\ref{SC-2}.
For the Hg system, $R$ is much more suppressed because the $d_{z^2}$-orbital based band is almost localized and detached from the $d_{x^2-y^2}$-orbital based band [see Fig.~\ref{band-DOS}(c)].
This is an ideal condition for superconductivity~\cite{Sakakibara} and therefore $P^{dd}$ becomes largest for the Hg system.

These results conclude that superconductivity is more enhanced when the $d_{z^2}$-orbital based band is deeply sinking and its contribution to the low-energy physics is small.
Therefore, the material dependence of superconductivity is understood only by incorporating the $d_{z^2}$ orbital explicitly into a model such as our model, which is a remarkable advantage over the usual one-band Hubbard and $t$-$J$ models, and even the three-band $d$-$p$ model.

$P^{dd}$ calculated here corresponds to the square of the superconducting order parameter and is closely related to the superconducting transition temperature $T_{\text{c}}$.
The critical doping rate $\delta_{\text{c}}$ for the La system, where $P^{dd}$ becomes zero, seems to be too small ($\sim0.16$) compared with the experimental value ($\delta_{\text{c}}=0.25-0.3$).
Furthermore, the sudden disappearance of $P^{dd}$ is also unrealistic.
It can be inferred that the actual value of $\varepsilon_{d_{z^2}}$ is much lower than the value estimated from the LDA calculation.
Indeed, as shown in Table~\ref{hopping}, the value of $\varepsilon_{d_{z^2}}$ with reference to the QSGW calculation is much lower.
Furthermore, the QSGW band structure~\cite{Jang1} can well explain the resonant inelastic X-ray scattering experiment~\cite{MorettiSala}.
We expect that the revised band structure shown in Fig.~\ref{band-DOS}(b) properly includes the correction for the LDA calculation and gives more realistic result for the La system.

\subsubsection{Material dependence: Effect of apical oxygen height}
From the viewpoint of the actual lattice structure, $\varepsilon_{d_{z^2}}$ is governed by the apical oxygen height, i.e., the distance between the apical oxygen and the copper:
The larger the apical oxygen height is, the lower $\varepsilon_{d_{z^2}}$ is with respect to $\varepsilon_{d_{x^2-y^2}}$, because of the crystal field effect~\cite{Kamimura,Kuroki}.
In addition, a larger apical oxygen height leads to a lower site energy $\varepsilon_{p_z}$ of the apical oxygen due to the decrease of a crystal field effect,
which in turn lowers the $\varepsilon_{d_{z^2}}$ through the hybridization between $p_z$ and $d_{z^2}$ orbitals despite the increase of the distance between apical oxygen and the copper.
Therefore, our result suggests that a larger apical oxygen height leads to a higher $T_{\text{c}}$ through a lower $\varepsilon_{d_{z^2}}$.
Indeed, the experimentally observed apical oxygen height of the Hg system is larger than that of the La system.
This tendency is also consistent with the so-called Maekawa's plot~\cite{Ohta}, where a lower $\varepsilon_{p_z}$ is related to a higher $T_{\text{c}}$.
Although the model itself does not explicitly include the $p_z$ orbital of the apical oxygen, the present four-band $d$-$p$ model properly incorporates the effect of the apical oxygen height via adjusting the site energy $\varepsilon_{d_{z^2}}$.

\subsubsection{Material dependence: Effect of $\Delta_{dp}$}
We also show in Fig.~\ref{Rz}(b) the quasiparticle renormalization factor $z$ estimated from the jumps in the total momentum distribution function $n^{\text{tot}}(\textbf{k})$ along the nodal direction of the $d_{x^2-y^2}$-wave superconducting gap.
At $\delta=0$, the system is insulating and thus $z=0$.
With increasing $\delta$, $z$ increases according to the decrease of the electron correlation effect.
We find that $z$ for the La system is smaller than that for the Hg system, indicating that the electron correlation effect is stronger in the La system.
This can be attributed to the larger $\Delta_{dp}=\varepsilon_{d_{x^2-y^2}}-\varepsilon_{p}$ for the La system, which results in the larger $d$-orbital occupancy of holes when it is doped.
The electron correlation has a dual effect:
One is to enhance the superconducting pairing and the other is to reduce the mobility of Cooper pairs.
In the present case, the latter effect would be dominant and thus $P^{dd}$ for the La system is suppressed compared with the Hg system.
This tendency is consistent with the negative correlation between $\Delta_{dp}$ and $T_{\text{c}}$ as mentioned in Sec.~\ref{intro}.

\begin{figure}
\centering
\includegraphics[width=0.7\hsize]{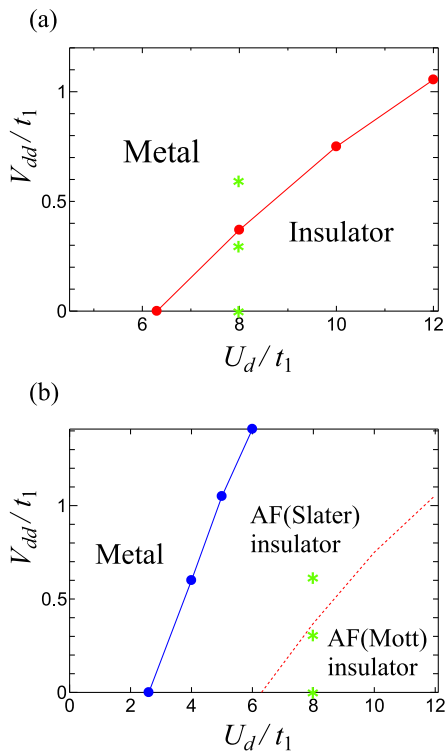}
\caption{\label{MIT}
Ground state phase diagram for the Hg system at $\delta=0$ (a) within the paramagnetic phase and (b) with the AF phase.
The dotted curve in (b) corresponds to the solid curve in (a).
Green stars in (a) and (b) represent the parameters used in Fig.~\ref{SC-3}.
}
\end{figure}

\subsection{Effect of $V_{dd}$}\label{Vdd}
Now we discuss the effect of the Coulomb interaction between nearest-neighbor $d$ orbitals, $V_{dd}$.
$V_{dd}$ is expected to be smaller than other Coulomb interactions, $U_d, U'_d, U_p, V_{dp}$, and $V_{pp}$~\cite{Hirayama}.
However, $V_{dd}$ directly affects the charge and spin correlations between nearest-neighbor $d$ electrons, which dominate the properties of cuprate superconductors.
As discussed in this section, we verify that the superconductivity in the model studied here is more sensitive to the value of $V_{dd}$ than other Coulomb interactions.
Hereafter, we treat only $U_d$ and $V_{dd}$ as independent parameters, and set other Coulomb interactions as $(U'_d,J,U_p,V_{dp},V_{pp})=(0.8, 0.1, 0.5, 0.25, 0.2)\;U_d$ with reference to Ref.~\cite{Hirayama}

\begin{figure}
\centering
\includegraphics[width=0.8\hsize]{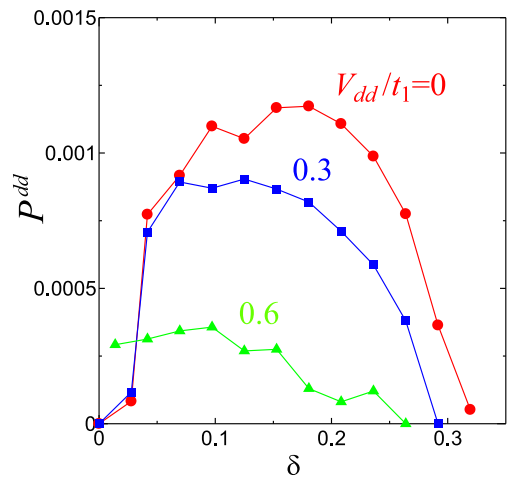}
\caption{\label{SC-3}
$P^{dd}$ as a function of the hole doping rate $\delta$ at $U_d/t_1=8$ and $V_{dd}/t_1=0, 0.3, 0.6$ for the Hg system.
}
\end{figure}

Figure~\ref{MIT}(a) shows the ground state phase diagram for the Hg system at $\delta=0$ within the paramagnetic phase.
When $V_{dd}=0$, a metal-insulator transition occurs at $U_d/t_1\sim6.3$, assuming the paramagnetic phase.
The metal-insulator transition is detected by monitoring the jump in the total momentum distribution function $n^{\text{tot}}(\textbf{k})$ as well as the long-range behavior of the charge Jastrow factor $P_{\text{J}_{\text{c}}}$~\cite{Watanabe2}.
This transition is a Mott metal-insulator transition because $\delta=0$ corresponds to the case with one hole per unit cell, $n_{\text{hole}}=1$, where the system cannot be a band insulator.
With the introduction of small but finite $V_{dd}$ (one order of magnitude smaller than $U_d$), the metallic region is substantially enlarged.
This is understood because the densities of empty sites and doubly-occupied sites increase to reduce the energy loss of $V_{dd}$, thus effectively weakening $U_d$ by $V_{dd}$, and then the insulating phase is destabilized.

It is noteworthy that the AF insulator is always more stable than the paramagnetic insulator in the present parameter space.
With increasing $U_d$, the system undergoes the phase transition from the metallic phase to a Slater-type AF insulator [a blue line in Fig.~\ref{MIT}(b)],
followed by the crossover to a Mott-type AF insulator [a red dotted line in Fig.~\ref{MIT}(b)].
Here, a Slater-type AF insulator is an insulator that becomes metallic without AF order, while a Mott-type AF insulator is an insulator that remains insulating without AF order.
The crossover line in Fig.~\ref{MIT}(b) is hence identical with the paramagnetic metal-insulator transition line in Fig.~\ref{MIT}(a).
As explained next, it is crucially important for the appearance of high-$T_{\text{c}}$ superconductivity whether the AF insulator in the parent compounds ($\delta$=0) is Slater-type or Mott-type~\cite{Weber2,Weber3,Jang2,Watanabe2}.

Next we study the effect of $V_{dd}$ on superconductivity.
Figure~\ref{SC-3} shows the superconducting correlation function $P^{dd}$ at $U_d/t_1=8$ and $V_{dd}/t_1=0,0.3,$ and 0.6 for the Hg system.
As $V_{dd}$ increases, $P^{dd}$ is suppressed and the peak position is moved to a smaller value of $\delta$.
In particular, a significant suppression is observed when $V_{dd}/t_1=0.6$.
In this case, $P^{dd}$ vs. $\delta$ does not show a dome-shaped behavior but instead an almost monotonic decrease, which rather reminds us of electron-doped cuprates~\cite{Matsumoto,Krockenberger}.
This observation of $P^{dd}$ vs. $\delta$ corresponds to the fact that the system is out of the Mott insulator region at $\delta=0$ (see Fig.~\ref{MIT}).
The view of a ``doped Mott insulator'' is thus no longer valid.
Similar claims have been made in the study of one-band Hubbard models~\cite{Yokoyama2,Tocchio}.
Our result suggests that it is essential to start from the Mott insulator region at $\delta=0$ to reproduce the dome-shaped behavior observed experimentally in hole-doped cuprates.
This can be considered as a good criterion for choosing the reasonable Coulomb interaction parameters of an effective model for cuprates.

\subsection{Phase competition}\label{competition}
Finally, we briefly discuss the competition between superconductivity and other phases.
The energy comparison among various phases in the ground state is a subtle problem and depends significantly on the numerical methods.
The VMC method used here tends to overestimate the magnetic long-range ordered phases, although it is much improved as compared with a mean-field type approximation.
In fact, we find that the AF phase and the stripe phase with both spin and charge modulations have lower variational energies than $d_{x^2-y^2}$-wave superconductivity for $\delta<0.3$.
Nevertheless, we believe that the present results capture the essence of the material dependence of cuprate superconductors and the conclusion is unchanged,
when the improved wave functions incorporating the quantum fluctuations suppress these overestimated competing orders.
We also note that the effect of $V_{dd}$ is also important for the phase competition.
This is because most of the competing phases including the phase separation are governed by the correlation between nearest-neighbor $d$ electrons.
This is also the case in the one-band Hubbard model~\cite{Misawa,Ohgoe}.
The detailed ground state phase diagram, including various competing phases, for the four-band $d$-$p$ model is left for a future study.

\section{DISCUSSION AND SUMMARY}\label{summary}
To obtain the unified description of cuprate superconductors, we have studied the four-band $d$-$p$ model for the La$_2$CuO$_4$ and HgBa$_2$CuO$_4$ systems.
We have shown that a lower $\varepsilon_{d_{z^2}}$ with respect to $\varepsilon_{d_{x^2-y^2}}$ and a smaller $\Delta_{dp}(>0)$ lead to a higher $T_{\text{c}}$.
The former results in a more localized $d_{z^2}$-orbital based band that do not interfere the superconductivity.
The latter results in a larger $z$, namely, a weaker electron correlation effect, which promotes the itinerancy of mobile carriers and thus enhances superconductivity.
The present four-band $d$-$p$ model covers these two factors, beyond the usual one-band and even three-band $d$-$p$ models.
Therefore, this model is considered to be a minimal model that can properly describe the material dependence of cuprate superconductors,
and thus it can also provide a valuable guideline to design new materials with a higher $T_{\text{c}}$.

The effect of $V_{dd}$ has also been investigated.
Although the value of $V_{dd}$ is small compared with other Coulomb interaction parameters, it substantially affects the ground state property of the system.
$V_{dd}$ weakens the effective $U_d$ and induces the paramagnetic metal-insulator transition, or the crossover from a Slater insulator to a Mott insulator.
The stability of superconductivity is also affected by $V_{dd}$.
Considering the doping dependence, we have to start from the Mott insulator region at $\delta=0$ to obtain the stable superconductivity and the dome-shaped dependence of $P^{dd}$ as a function of $\delta$.
Therefore, the appropriate estimation of $V_{dd}$ is important for the modelling of cuprates, as in other strongly-correlated electron systems where various phases compete.

\begin{acknowledgments}
The authors thank K. Kuroki for useful discussions.
The computation has been done using the facilities of the Supercomputer Center, Institute for Solid State Physics, University of Tokyo and the supercomputer system HOKUSAI in RIKEN.
This work was supported by a Grant-in-Aid for Scientific Research on Innovative Areas ``Quantum Liquid Crystals'' (KAKENHI Grant No. JP19H05825) from JSPS of Japan,
and also supported by JSPS KAKENHI (Grant Nos. JP21H04446, JP20K03847, JP19K23433, JP19H01842, and JP18H01183).

\end{acknowledgments}

\appendix*
\section{Construction of the trial wave function}
Here, we describe the details of the trial wave function together with the noninteracting tight-binding model obtained on the basis of the first-principles calculations in Sec.~\ref{band}.
The construction of the trial wave function is the most important part for the VMC method.
Depending on the trial state, both real and $\textbf{k}$ space representations are used.

\subsection{Noninteracting energy band}\label{NIband}
First, we describe how to construct the noninteracting tight-binding energy band discussed in Sec.~\ref{band}.
%The one-body part $\left|\Phi\right>$ can be 
The noninteracting energy band is obtained by diagonalizing the following one-body Hamiltonian:
\begin{widetext}
\begin{align}
H_{\text{kin}}&=\sum_{\textbf{k},\sigma}
        \left( c_{\textbf{k}1\sigma}^{\dg}, c_{\textbf{k}2\sigma}^{\dg},  
   c_{\textbf{k}3\sigma}^{\dg}, c_{\textbf{k}4\sigma}^{\dg}\right) 
 \begin{pmatrix}
   t_{11} & t^*_{21} & t^*_{31} & t^*_{41} \\ 
   t_{21} & t_{22} & t^*_{32} & t^*_{42} \\
   t_{31} & t_{32} & t_{33} & t^*_{43} \\
   t_{41} & t_{42} & t_{43} & t_{44}
 \end{pmatrix}
 \begin{pmatrix}
  c_{\textbf{k}1\sigma} \\ c_{\textbf{k}2\sigma} \\ 
  c_{\textbf{k}3\sigma} \\ c_{\textbf{k}4\sigma}    
 \end{pmatrix} \label{kin_band_app1}\\
      &=\sum_{\textbf{k},\sigma}\sum_m E_m(\textbf{k})a^{\dg}_{\textbf{k}m\sigma}a_{\textbf{k}m\sigma} \label{kin_band_app2}
\end{align}
with the hopping matrix elements given as
\begin{align}
t_{11}&=\varepsilon_{d_{x^2-y^2}},\\
t_{21}&=0,\\
t_{22}&=\varepsilon_{d_{z^2}}-2t_5(\cos k_x+\cos k_y),\\
t_{31}&=2\text{i}t_1\sin\frac{1}{2}k_x,\\
t_{32}&=-2\text{i}t_4\sin\frac{1}{2}k_x,\\
t_{33}&=\varepsilon_{p_x}+2t_3\cos k_x+2t_6[\cos(k_x+k_y)+\cos(k_x-k_y)],\\
t_{41}&=-2\text{i}t_1\sin\frac{1}{2}k_y,\\
t_{42}&=-2\text{i}t_4\sin\frac{1}{2}k_y,\\
t_{43}&=2t_2\left[\cos\left(\frac{1}{2}k_x+\frac{1}{2}k_y\right)-\cos\left(\frac{1}{2}k_x-\frac{1}{2}k_y\right)\right],\\
t_{44}&=\varepsilon_{p_y}+2t_3\cos k_y+2t_6[\cos(k_x+k_y)+\cos(k_x-k_y)],
\end{align}
\end{widetext}
where $c^{\dg}_{\textbf{k}\alpha\sigma}$ ($c_{\textbf{k}\alpha\sigma}$) is a creation (annihilation) operator of an electron with momentum $\textbf{k}$, spin $\sigma\,(=\ua,\da)$, and orbital $\alpha\, (=1,2,3,4)$
corresponding to ($d_{x^2-y^2}$, $d_{z^2}$, $p_x$, $p_y$), respectively.
The bonds between nearest-neighbor Cu sites are set as unit vectors ($|\textbf{e}_x|=|\textbf{e}_y|=1$) and the bonds between nearest-neighbor Cu and O sites are one-half of them (see Fig.~\ref{lattice}).
The hopping integrals $t_i$ ($i=1-6$) and the site energy of each orbital $\varepsilon_{\alpha}$ are determined by fitting the band structures that are obtained by the LDA or QSGW calculation.
The specific values for the La, La(revised), and Hg systems are listed in Table~\ref{hopping}.
Equation (\ref{kin_band_app2}) is obtained by diagonalizing the Hamiltonian matrix in Eq.~(\ref{kin_band_app1}) and is the same with Eq.~(\ref{kin_band}) in Sec.~\ref{d-p}.
$E_m(\textbf{k})$ is the noninteracting energy band characterized by the wave vector $\textbf{k}$ and the energy band index $m\,(=1,2,3,4)$ with $a^{\dg}_{\textbf{k}m\sigma}$ ($a_{\textbf{k}m\sigma}$)
being a creation (annihilation) operator of the corresponding energy band with spin $\sigma$.

\subsection{Trial wave function}
\subsubsection{Superconductivity}
To construct the trial wave function for superconductivity, we employ the Bogoliubov de-Gennes (BdG) type Hamiltonian in real space~\cite{Himeda}, i.e., 
\begin{equation}
H_{\text{BdG}}=\sum_{i,j}\sum_{\alpha,\beta}\left(c^{\dg}_{i\alpha\ua}, c_{i\alpha\da}\right)
 \begin{pmatrix}
  T^{\alpha\beta}_{ij\ua} & \Delta^{\alpha\beta}_{ij} \\ 
  \Delta^{\alpha\beta*}_{ji} & -T^{\alpha\beta}_{ji\da}
 \end{pmatrix}
 \begin{pmatrix}
  c_{j\beta\ua} \\
  c^{\dg}_{j\beta\da}
 \end{pmatrix}.
\end{equation}
Here, $T^{\alpha\beta}_{ij\sigma}$ is obtained from the Fourier transform of the matrix in Eq. (\ref{kin_band_app1}) with renormalized hopping integrals $\tilde{t}_i$ and also includes the chemical potential term.
The chemical potential $\mu$ is set to the Fermi energy in the noninteracting limit.
$\Delta^{\alpha\beta}_{ij}$ corresponds to an anomalous part that describes the superconducting pairing in real space.
Therefore, the variational parameters to be optimized in $\left|\Phi\right>$ are $\tilde{t}_i$ ($i=2-6$) and $\{\Delta^{\alpha\beta}_{ij}\}$ with $\tilde{t}_1=t_1$ being fixed as a unit of energy.
In this study, we consider the pairing between nearest-neighbor orbitals, $d$-$d$, $d$-$p_x$, $d$-$p_y$, $p_x$-$p_x$, and $p_y$-$p_y$, where $d$ denotes the $d_{x^2-y^2}$ orbital.
In the paramagnetic phase, we simply set $\Delta^{\alpha\beta}_{ij}=0$.

We can also construct the trial wave function with the band ($\textbf{k}$ space) representation,
where $\Delta^{mn}(\textbf{k})\propto \bigl< a_{\textbf{k}m\ua}a_{-\textbf{k}n\da}\bigr>$ is the variational parameter.
However, we find that the trial wave function with the real space representation always gives lower (i.e., better) energy than that with the band representation, especially, for large Coulomb interaction parameters.
This is because the Coulomb interaction depends on the orbital, not on the band, and thus the trial wave function with the real space (orbital) representation gives the better result.

\subsubsection{Uniform spin AF and stripe phases}
As mentioned in Sec.~\ref{VMC}, various long-range orderings of charge and spin can be described by introducing $\{\rho^{\alpha}_i\}$ and $\{m^{\alpha}_i\}$.
A uniform spin AF phase with A and B sublattices for the orbital $\alpha$ is expressed with the staggered potential
\begin{equation}
m^{\alpha}_i=
\begin{cases}
-s_{\sigma} m^{\alpha} & \text{for A sublattice} \\
+s_{\sigma} m^{\alpha} & \text{for B sublattice}
\end{cases}
\end{equation}
where $s_{\sigma}=1(-1)$ for $\sigma=\ua(\da)$.
For a stripe phase with charge and spin periodicities $\lambda^{\alpha}_{\text{c}}=2\pi/q^{\alpha}_{\text{c}}$ and $\lambda^{\alpha}_{\text{s}}=2\pi/q^{\alpha}_{\text{s}}$, respectively, the following potentials with spatial modulation in the $x$ direction should be introduced:
\begin{equation}
\rho^{\alpha}_i=\rho^{\alpha}\cos[q^{\alpha}_{\text{c}}(x_i-x^{\alpha}_{\text{c}})]
\end{equation}
and
\begin{equation}
m^{\alpha}_i=(-1)^{x_i+y_i}m^{\alpha}\sin[q^{\alpha}_{\text{s}}(x_i-x^{\alpha}_{\text{s}})],
\end{equation}
where $x^{\alpha}_{\text{c}}$ and $x^{\alpha}_{\text{s}}$ control the relative phases of charge and spin orderings, respectively.
For example, the stripe phase observed around $\delta=1/8$ in several cuprate superconductors corresponds to $\lambda_{\text{c}}=4$ and $\lambda_{\text{s}}=8$, although the orbital dependence and relative phase are still under debate.

\end{document}